\documentstyle[preprint,aps,epsfig]{revtex}
\title{Spin-Orbit-Induced Magnetic
Anisotropy for Impurities in Metallic Samples I. Surface Anisotropy}
\author{O. \'Ujs\'aghy$^a$ and A. Zawadowski$^{a,b}$} 
\address{$^a$Institute of Physics and Research Group of Hungarian Academy
of Sciences, Technical University of Budapest,
H-1521 Budapest, Hungary}
\address{$^b$Research Institute for Solid State Physics, POB 49, H-1525
Budapest, Hungary}
\date{\today}
\begin{document}
\draft
\maketitle

\begin{abstract}
Motivated by the recent measurements of Kondo resistivity in thin films
and wires, where the Kondo amplitude is suppressed for thinner samples,
the surface anisotropy for magnetic impurities is studied.
That anisotropy is developed in those cases where in addition to the 
exchange interaction with the impurity there is strong spin-orbit 
interaction for conduction electrons around the impurity in the
ballistic region.
The asymmetry in the neighborhood of the magnetic impurity exhibits the
anisotropy axis ${\bf n}$ which, in the case of a plane surface, is 
perpendicular to the surface.
The anisotropy energy is $\Delta E=K_d ({\bf n}{\bf S})^2$ for spin ${\bf S}$,
and the anisotropy constant $K_d$ is inversionally proportional to 
distance $d$ measured from the surface and $K_d>0$. Thus at low temperature 
the spin is frozen in a singlet or doublet of lowest energy. 
The influence of that anisotropy on the electrical resistivity is the
subject of the following paper (part II).
\end{abstract}


\section{Introduction}
\label{sec:1}

In the last few years the experimental study of dilute magnetic alloys 
with non-magnetic host of reduced dimensions has attracted considerable 
interest \cite{Blachy,Chandra,Giordano}. The subjects of most of the 
experimental works \cite{Blachy,Chandra} are the size dependence of the 
Kondo effect, thus to determine whether the Kondo temperature and 
amplitude depend on the film thickness or the diameter of the wire or not.
A very recent paper of N. Giordano \cite{Giordano} has indicated that the 
magnetoresistance above  the Kondo temperature depends also 
on the thickness of the film.

\begin{figure}
  \begin{center}
    \epsfig{file=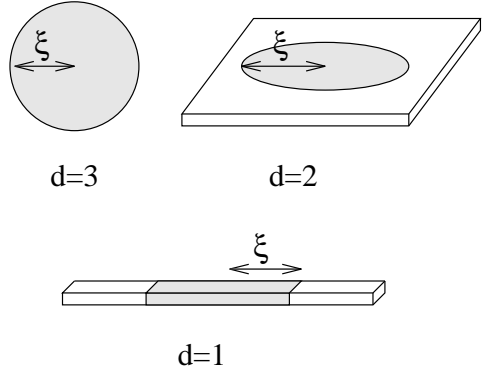}
  \end{center}
\caption{The Kondo sreening cloud in different dimensions.}
\label{figfelho}
\end{figure}

The theoretical motivation of these experiments has been the concept of spin 
compensated Kondo state.
Considering the Kondo effect the ground state is a singlet where the spin 
of the magnetic impurity is screened by the spin polarization of the conduction
electrons, which is known as the screening or compensation cloud \cite{Wilson}.
The theoretical studies of the Kondo effect suggest, that the size of that 
screening cloud, $\xi$ is of the order of $\hbar v_F/(k_B T_K)$ where 
$v_F$ is the
Fermi velocity and $T_K$ is the Kondo temperature \cite{Affleck}.
That length scale is especially large for alloys with small Kondo 
temperature thus for Au(Fe) it can be in range of $\xi=10^5\text{\AA}$
with $T_K\sim 1$ K. In the case of wires or films it is easy to prepare 
such samples where the size at least in one direction is smaller than that
Kondo coherence length $\xi$. The question has been raised concerning those 
experiments where the size is smaller, whether the coherence length 
does prevent 
the formation of the spin-compensated ground state or not 
\cite{Blachy,Bergmann}. Even if that argument looks very challenging, the 
theoretical base for that argument is very weak, as the magnetic impurity 
experiences the conduction electron density only at the site of the impurity.
At zero temperature in case $S=1/2$ the polarization cloud must contain one 
electron with spin antiparallel to the local spin. The decay of the cloud
is determined by the correlation function 
$\langle {\bf S}\bbox{\sigma}({\bf r})\rangle$ 
where ${\bf S}$ stands for the impurity spin located at
${\bf r}=0$ and $\bbox{\sigma}({\bf r})$ is the spin polarization of the
conduction electron. In dimension $d=3$ the correlation 
$\langle {\bf S}\bbox{\sigma}
({\bf r})\rangle$ decays like $r^{-2}$, but in reduced dimensions the
decay is weaker, for $d=2$ ($d=1$) it is like $r^{-1}$ ($r^0$), the
Kondo coherence length is, however, not affected \cite{Affleck}.
Beyond the distance $\xi$ the decay is exponential like $\exp\{-r/\xi\}$.
Thus in $d=3$ the shape of the cloud is sphere, in $d=2$ pancake, and
in $d=1$ cigar like (see Fig.\ \ref{figfelho}). Following that argument
only the level spacing $\Delta E$ can hinder the formation of the 
compensated groundstate if $\Delta E\geq k_B T_K$. That can happen only for 
grains of size about $100\text{\AA}$, but that is out of question for films and
wires if the electrons are not localized \cite{Bergmann}.
Thus accepting the existence of that size dependence an other explanation is 
required, but there are also experiments where the existence of the size 
dependence is questioned \cite{Chandra}.

The influence of non-magnetic impurities on the formation of 
the Kondo resonance has been investigated for thin film using
the theory of weak localization \cite{Martin}.
In contrast to that, the present work deals with the ballistic region.

Recently it has been suggested by B. L. Gyorffy and the authors of the 
present paper \cite{Ujsaghy} that for the magnetic impurity interacting with
the conduction electrons by the effective exchange interaction 
a surface magnetic anisotropy can be the result of
spin-orbit scattering of the conduction electrons on the non-magnetic host.
The magnetic anisotropy energy is given for a single impurity by the formula
\begin{equation}
H_a=K_d ({\bf n}{\bf S})^2,
\label{Haniz}\end{equation}
where $({\bf n}{\bf S})$ is the spin component of the impurity spin in the 
direction parallel to the normal vector ${\bf n}$ of the surface and the
amplitude $K_d>0$ is inversely proportional to the distance of the impurity
$d$, measured from the surface.

The magnetic anisotropies caused by the relativistic corrections occuring 
in the Dirac equation, as the dipole-dipole and the spin-orbit interactions 
can reflect the geometry of the sample. For example in magnets they are 
responsible for the easy axis magnetization where the dipole-dipole
term is dominating \cite{Van Vleck}. In the case of superimposed magnetic 
and non-magnetic layers a magnetic anisotropy is developed 
from first principles which is formed 
as a result of competition between those two interactions 
\cite{Szunyogh1,Guo}.

As far as it is known by the present authors until recently
the possibility has not been explored that the 
spin-orbit interaction between the non-magnetic host atoms and the
conduction electrons can produce a magnetic anisotropy for the magnetic
impurities by the exchange interaction between the impurity spin and
the conduction electrons. Such an anisotropy cannot develop for the 
impurity in the bulk, but that can exist in host
limited in space. Thus, that anisotropy reflects the geometry of the sample
and the position of the impurity in that.

The present paper is devoted to calculate that anisotropy in the second order
both in the exchange interaction between the impurity and the electrons and
in the spin-orbit interaction between the electrons and the host atoms. The 
mean field calculation does not lead to such terms.

The present paper is organized as follows. In Sec.~\ref{sec:2} the model 
is described where it is assumed that the spin-orbit interaction takes place 
on the localized e.g. d-levels of the host atoms and that localized orbital 
is hybridized with the conduction electrons following the idea of the
Anderson model \cite{Anderson}. In Sec.~\ref{sec:3} the electron Green's 
function is calculated 
in the first order of the spin-orbit interaction. Sec.~\ref{sec:4} is devoted
to calculate the impurity spin self-energy in which the spin anisotropy given
by Eq.\ (\ref{Haniz}) shows up. The expression for the anisotropy constant is 
developed in Sec.~\ref{sec:5} \cite{proc}. The conclusion is presented 
in Sec.~\ref{sec:6}.
In Appendix~\ref{app:1} the role of the time reversal symmetry is explored.
The complicated integrals appearing in the final expression of the anisotropy 
constant are developed by analytical calculations in Appendix~\ref{app:2}.
The following paper, Part II \cite{Ujsaghy2} deals with 
the calculation of the amplitude of 
the Kondo resistivity anisotropy as a function of the film thickness
\cite{proc}. 
The study of the magnetoresistance is left for a further publication
\cite{Borda}. 
The comparison with the experiments is contained by Part II
\cite{Ujsaghy2}.

\section{The model}
\label{sec:2}

\begin{figure}
  \begin{center}
    \epsfig{file=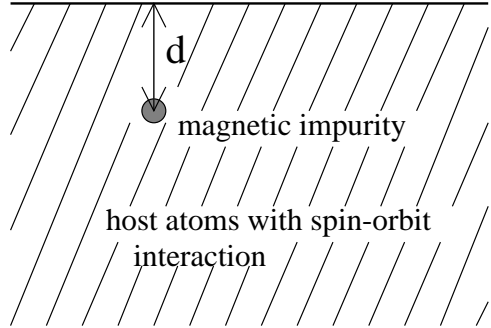}
  \end{center}
\caption{The infinite half space of homogeneously dispersed
host atoms with spin-orbit interaction and the magnetic
impurity in a distance $d$ measured from the surface.}
\label{fig1}
\end{figure}

For the sake of simplicity we consider an infinite half space where the
host atoms with spin-orbit interaction are homogeneously dispersed (no
crystal structure effect), and the magnetic impurity is placed in a 
distance $d$ from
the surface (see Fig.\ \ref{fig1}). For further simplification the shape of 
the sample is taken into account only in the positions of the host atoms
representing the spin-orbit interaction, while the 
free electron like conduction electrons move in the unlimited whole space.

The interaction between the conduction electrons and the magnetic impurity
is described by the simplest realistic Hamiltonian with orbital quantum
numbers. Therefore $S=5/2$ is chosen, because in this case the Hamiltonian
is diagonal in the orbital quantum number $m$ according to the Hund's rule.
For other values of $S$ the Hamiltonian consists of several complicated 
terms \cite{Nozieres,Muhl}. After these considerations we can 
write the Hamiltonian as 
\begin{eqnarray}
H_0&=&\sum\limits_{k,m,\sigma}\!\!\varepsilon_k \,a_{km\sigma}^\dagger
a_{km\sigma} \nonumber \\
&+& J\sum\limits_{\scriptstyle k,k^\prime,m,m^\prime \atop
\scriptstyle \sigma,\sigma^\prime}
\bbox{S}\,(a_{km\sigma}^\dagger \bbox{\sigma}_{\sigma\sigma'}
a_{k'm'\sigma'})\,\delta_{mm'}
\label{H0}\end{eqnarray}
where $a_{klm\sigma}^\dagger$ ($a_{klm\sigma}$) creates (annihilates)
an electron with momentum $k$, angular momentum $l,m$ and spin
$\sigma$, $J$ is the effective Kondo coupling, $\bbox{\sigma}$ stands for
the Pauli matrices and the origin is placed at the impurity site. Keeping
only the $l=2$ channels the index $l$ is dropped.

In order to make transparent that the spin-orbit interaction 
is mainly in the d-channel and due to the Coulomb potential of the 
nuclei we introduce a simple model where the spin-orbit interaction takes 
place on the d-levels of the host, which hybridizes with the conduction 
electrons.
These host atom orbitals are labeled by $n$ referring to the position
${\bf R}_n$ and also by the quantum numbers $l,m,\sigma$ (e.g. $l=2$ for
Cu and Au host and the index $l$ is dropped again).
The Hamiltonian of these extra orbitals is
\begin{eqnarray}
H_1&=&\varepsilon_0\sum\limits_{nm\sigma}\!\!b^{(n)\dagger}_{m\sigma}
b^{(n)}_{m\sigma}
+\lambda\sum\limits_{\scriptstyle nmm' \atop \scriptstyle
\sigma\sigma'}\!\!\langle m|
\bbox{L}|m'\rangle \langle\sigma|\bbox{\sigma}|\sigma'\rangle
b^{(n)\dagger}_{m\sigma} b^{(n)}_{m'\sigma'} \nonumber \\
&+&\sum\limits_{nk mm^\prime \sigma}\!\!\biggl(
V_{kmm'}({\bf R}_n)\,b^{(n)\dagger}_{m\sigma} a_{km'\sigma} + {\rm h.c.}
\biggr)
\end{eqnarray}
where $ b^{(n)\dagger}_{m\sigma}$ ($b^{(n)}_{m\sigma}$) creates
(annihilates) the host atom orbital at site $n$
with wave function 
$\Psi^{(n)}_{l=2,m}$. 
$V_{kmm'}({\bf R}_n)$ is the
Anderson' hybridization matrix element \cite{Anderson}, which depends 
on ${\bf R}_n$ since
spherical wave representation with origin at the magnetic impurity is
used, $\lambda$ is the strength of the spin-orbit coupling, and {\bf L} is 
the orbital momentum at site $n$. 
As the spin-orbit interaction is weak, therefore, its effect will 
be considered as a perturbation.
 
The ${\bf R}_n$ dependence in the Anderson' hybridization matrix element
can be evaluated as
\begin{equation}
V_{kll'mm'}({\bf R}_n)=\langle\Psi^{(n)}_{lm}|H_{eff}|\Phi^0_{kl'm'}\rangle,
\label{V1}\end{equation}
where $H_{eff}$ is the effective hybridization Hamiltonian between the host
atom orbital ($\Psi^{(n)}_{lm}$) and the conduction electron states with
origin at the magnetic impurity ($\Phi^0_{kl'm'}$). After inserting a 
complete orthonormal set of free spherical waves with origin at the
host atom at ${\bf R}_n$
\begin{equation}
\int\limits_0^{\infty} dk^{''}\sum\limits_{l^{''},m^{''}} |
\Phi^{(n)}_{k^{''}l^{''}m^{''}}\rangle\langle
\Phi^{(n)}_{k^{''}l^{''}m^{''}}|=1,
\label{ONR}\end{equation}
and taking into account the usual assumption 
\begin{equation}
\langle\Psi^{(n)}_{lm}|H_{eff}|\Phi^{(n)}_{k''l''m''}\rangle=
V_{l}\delta_{l''l}\delta_{m''m}
\label{felt}\end{equation}
so that the hybridization matrix element is diagonal in quantum numbers 
$l$, $m$, shows slow $k$-dependence and it is the same for each
host atom we got for the hybridization matrix element
\begin{equation}
V_{kll'mm'}({\bf R}_n)=V_{l}\int\limits_0^{\infty} dk''
\langle\Phi^{(n)}_{k''lm}|
\Phi^0_{kl'm'}\rangle.
\label{V2}\end{equation}
Thus to get the hybridization matrix element one has to calculate the overlap 
between spherical waves with different origin,
\begin{equation}
\langle\Phi^{(n)}_{k^{''}lm}|\Phi^0_{kl'm'}\rangle=\int d{\bf r}
\Phi^*_{k^{''}lm}({\bf r})\Phi_{kl'm'}({\bf r}+{\bf R}_n)
\end{equation}
where $\Phi_{klm}({\bf r})=\sqrt{{2 k^2\over\pi}} j_l(k r) Y_l^m({{\bf r}
\over r})$ are free spherical waves, $j_l(k r)$ are spherical Bessel 
functions, 
$Y_l^m({\bf r})$ are spherical harmonics, and $r=|{\bf r}|$ \cite{Tannoudji}.
This can be simplified by
using a local coordinate system for each host atom  where the $z_{(n)}$
axis is directed parallel to ${\bf R}_n$. 
In that system $m$ is conserved and  
\begin{equation}
V_{kll'mm'}(R_n)=V_l\delta_{mm'}v_{kll'm}(R_n),
\label{V3}\end{equation}
where 
\begin{eqnarray}
v_{kll'm}(R_n)&=&S_l {2\over\pi}\sqrt{{(2 l+1)(l-m)!\over 2 (l+m)!}}
\sqrt{{(2 l'+1)(l'-m)!\over 2 (l'+m)!}} k R_n\nonumber \\
&&\cdot\int\limits_0^{\infty}
dy\int\limits_{|1-y|}^{1+y} dy' {y'\over y} j_{l'}(k y' R_n)
P_l^m\biggl({y'^2-y^2-1\over 2 y}\biggr) 
P_{l'}^m\biggl({y'^2-y^2+1\over 2 y'}\biggr),
\label{valt}
\end{eqnarray}
and
\begin{equation}
S_l=\int\limits_0^{\infty} dx x j_l(x). 
\label{S}\end{equation}
In Eq.\ (\ref{valt}) the $y=r/R_n$, $y'=|{\bf r}+{\bf R_n}|/R_n$, 
in Eq.\ (\ref{S}) the $x=k''r$ new integration variables have been introduced, 
$P_l^m$ are Legendre polynomials, and $R_n=|{\bf R}_n|$.
The occuring integrals could be evaluated analytically.

For $l=l'=2$ $S_2=2$, thus by introducing the notation $V=V_2$
\begin{equation}
V_{kmm'}=V_{k22mm'}=V\delta_{mm'} v_{km}(R_n)
\label{V22}\end{equation}
where the $v_{km}(R_n)$ matrix elements are symmetric 
for $\pm m$ and show different power behaviors in $R_n$ at $k=k_F$:

\begin{mathletters}
\label{vkm}
\begin{equation}
v_{k_F0}(R_n)=10\biggl({\sin(k_F R_n)\over 2 k_F R_n}
+{3\cos(k_F R_n)\over (k_F R_n)^2}-{12\sin(k_F R_n)\over (k_F R_n)^3}
-{27\cos(k_F R_n)\over (k_F R_n)^4}+{27\sin(k_F R_n)\over (k_F R_n)^5}
\biggr),
\end{equation}
\begin{equation}
v_{k_F1}(R_n)=15\biggl( -{\cos(k_F R_n)\over (k_F R_n)^2}
+{5\sin(k_F R_n)\over (k_F R_n)^3}+{12\cos(k_F R_n)\over (k_F R_n)^4}
-{12\sin(k_F R_n)\over (k_F R_n)^5}\biggr),
\end{equation}
\begin{equation}
v_{k_F2}(R_n)=15\biggl( -{\sin(k_F R_n)\over (k_F R_n)^3}
-{3\cos(k_F R_n)\over (k_F R_n)^4}+{3\sin(k_F R_n)\over (k_F R_n)^5}
\biggr).
\end{equation}
\end{mathletters}

\section{Electron propagator in first order of spin-orbit coupling}
\label{sec:3}

\begin{figure}
  \begin{center}
    \epsfig{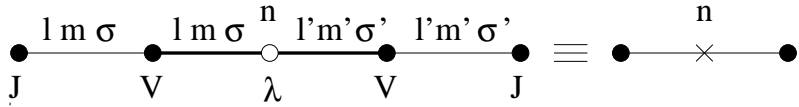}
  \end{center}
\caption{The electron
propagator leaving and arriving at the impurity.
The heavy lines represent the localized d-electron propagators, and
$V$ and $\lambda$ indicate the hybridization with the localized orbital and
the spin-orbit interaction, respectively. The indices are according to
the local system.}
\label{fig2}
\end{figure}

The electron propagator leaving and arriving at the impurity was calculated
in first order of spin-orbit coupling according to the diagram shown 
in Fig.\ \ref{fig2}.

In the local system the conduction electron propagator has the 
following matrix form in first order of spin-orbit coupling
\begin{equation}
G_{km\sigma,k'm'\sigma'}(i\omega_n)={\delta_{kk'}\delta_{mm'}
\delta_{\sigma\sigma'}\over i\omega_n-\varepsilon_k}+
\sum\limits_{n}
{1\over i\omega-\varepsilon_k}
W_{mm'\sigma\sigma'}(R_n){1\over i\omega-\varepsilon_{k'}}
\label{elprop}
\end{equation}
where
\begin{equation}
W_{mm'\sigma\sigma'}(R_n)=
V_{k_Fm\tilde m} G_d(\omega)\lambda\langle\tilde m|{\bf L}|\tilde m'\rangle
\langle\sigma|\bbox{\sigma}|\sigma'\rangle G_d(\omega) V^*_{k_F\tilde m'
m'}.
\label{Weloszor}
\end{equation}
The scatterings on several host atoms give higher order corrections
in $\lambda$.

In Eq.\ (\ref{Weloszor}) the $k$-dependence was replaced by $k_F$ and 
$G_d(\omega)$ denotes the electron propagator for the d-levels of the host 
atom. $G_d$ is given by the spectral function
$\rho_d$ as
\begin{equation}
G_d(\omega)=\int {\rho_d(\omega')\over \omega-\omega'} d\omega'
\end{equation}
where
\begin{equation}
\rho_d={1\over\pi}{\Delta\over (\omega-\varepsilon_d)^2+\Delta^2}
\end{equation}
and $\Delta=\pi V^2\rho_0$ is the width of the d-levels due to the
hybridization \cite{Anderson}, $\rho_0$ is the density of states
of the conduction electrons for one spin direction.
For $\omega\ll \max\{\varepsilon_d,\Delta\}$ $G_d$ can be replaced by a 
constant $1/\varepsilon_0$.
 
Thus using Eq.\ (\ref{V22})
\begin{equation}
W_{mm'\sigma\sigma'}(R_n)=
{\lambda V^2\over\varepsilon_0^2}
\biggl(B^+\sigma^-+B^-\sigma^++B^z\sigma^z\biggr)_{mm'\sigma\sigma'}
\label{W}
\end{equation}
where
$B^\pm$ and $B^z$ are 5$\times$5 matrices in the quantum number $m$,
having the form

\begin{mathletters}
\begin{equation}
B^+_{mm'}=\sqrt{(3+m')(2-m')} v_{k_Fm} v_{k_Fm'} \delta_{m,m'+1},
\end{equation}
\begin{equation}
B^-_{mm'}=\sqrt{(3-m')(2+m')} v_{k_Fm} v_{k_Fm'} \delta_{m,m'-1},
\end{equation}
\begin{equation}
B^z_{mm'}=m v_{k_Fm} v_{k_Fm'} \delta_{m,m'}.
\end{equation}
\end{mathletters}
These matrices could be introduced phenomenologically also.

By rotating back the local system to the frame of the sample where the
$z$ axis is perpendicular to the surface the electron propagator can 
be calculated. These rotations were done in the standard way by
using the formula
\begin{eqnarray}
&&\tilde W_{mm'\sigma\sigma'}(R_n,\theta_n,\varphi_n)=\nonumber\\
&& R^{(2)}_{m\bar m}(\varphi_n,\theta_n,0)R^{(1/2)}_{\sigma\bar\sigma}
(\varphi_n,\theta_n,0)W_{\bar m\bar m'\bar\sigma\bar\sigma'}(R_n)
R^{(2)}_{\bar m'm'}(0,-\theta_n,-\varphi_n)R^{(1/2)}_{\bar\sigma'\sigma'}
(0,-\theta_n,-\varphi_n),
\end{eqnarray}
where $(R_n,\,\theta_n,\,\varphi_n)$ are the polar coordinates of the
host atom labeled by $n$ in the system of the sample and 
$R^{(2)},\, R^{(1/2)}$ are the rotation matrices with angular 
momentum $J=2$ and $1/2$, respectively \cite{Brink}.
In a case when rotation symmetry around the $z$ axis of the system of the
sample is obeyed, as in the model described in Section~\ref{sec:2}, 
the electron propagator does not depend on the azimuthal angle $\varphi_n$ 
and thus it can be written as
\begin{equation}
\tilde W_{mm'\sigma\sigma'}(R_n,\theta_n)=
\delta_{m+\sigma,m'+\sigma'}
d^{(2)}_{m\bar m}(\theta_n)d^{(1/2)}_{\sigma\bar\sigma}
(\theta_n)W_{\bar m\bar m'\bar\sigma\bar\sigma'}(R_n)
d^{(2)}_{\bar m'm'}(-\theta_n)d^{(1/2)}_{\bar\sigma'\sigma'}
(-\theta_n)
\end{equation}  
where the Wigner-formula for rotation matrices \cite{Brink} was used.

The time reversal symmetry gives restrictions for the electron propagator
(see Appendix~\ref{app:1})
which provide a check of calculations. 
In the calculation the angular dependences are very important because in 
the case of s-wave scattering the spin-orbit interaction cannot
influence the dynamics of the impurity spin \cite{Meir}. 

\section{Self-energy corrections for the impurity spin}
\label{sec:4}

The self-energy was calculated by 
using Abrikosov's pseudofermion representation \cite{Abrikosov} for the 
impurity spin and Matsubara's diagram technique 
applied for the exchange interaction with coupling strength $J$ given
by Eq.\ (\ref{H0}). It can be shown that the Hartree-Fock diagram 
gives no contribution.

\begin{figure}
  \begin{center}
    \epsfig{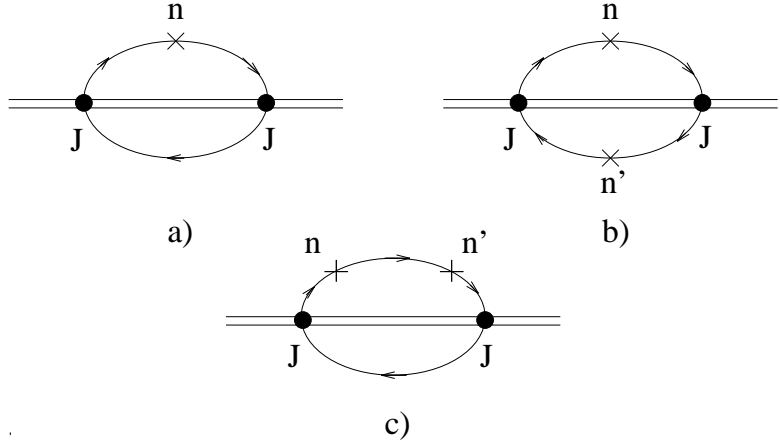}
  \end{center}
\caption{The self-energy diagrams for the impurity spin. The
double line represents the spin, the single one the conduction
electrons. The solid circles stand for the exchange interaction and the
$\times$ labelled by $n$ for the effective spin-orbit interaction on
the orbital of the host atom at ${\bf R}_n$.}
\label{fig3}
\end{figure}

The diagrams for the self-energy of the impurity spin which contain
the electron propagator calculated in Section~\ref{sec:3} are shown in
Fig.\ \ref{fig3}. 
The spin factors of these diagrams are 
\begin{mathletters}
\begin{equation}
J^2{\bf S}_{MM'}\bbox{\sigma}_{\sigma_3\sigma_1}\delta_{m_3m_1}
\tilde W_{m_1m_2\sigma_1\sigma_2}(R_n,\theta_n)
{\bf S}_{M'M}\bbox{\sigma}_{\sigma_2\sigma_3}\delta_{m_2m_3}
\end{equation}
\begin{equation}
J^2{\bf S}_{MM'}\bbox{\sigma}_{\sigma_3\sigma_1}\delta_{m_3m_1}
\tilde W_{m_1m_2\sigma_1\sigma_2}(R_n,\theta_n)
{\bf S}_{M'M}\bbox{\sigma}_{\sigma_2\sigma_4}\delta_{m_2m_4}
\tilde W_{m_4m_3\sigma_4\sigma_3}(R_{n'},\theta_{n'})
\end{equation}
\begin{equation}
J^2{\bf S}_{MM'}\bbox{\sigma}_{\sigma_4\sigma_1}\delta_{m_4m_1}
\tilde W_{m_1m_2\sigma_1\sigma_2}(R_n,\theta_n)
\tilde W_{m_2m_3\sigma_2\sigma_3}(R_{n'},\theta_{n'})
{\bf S}_{M'M}\bbox{\sigma}_{\sigma_3\sigma_4}\delta_{m_3m_4}
\end{equation}
\end{mathletters}
for a), b) and c), respectively. In the spin factor of diagram a) the 
trace of $\tilde W$ disappears. This easily can be seen
from the form of $W$ in the local coordinate system because $B^{\pm}$
and $B^z$ are traceless and trace is invariant under rotation.
The spin factor of diagram c) is proportional to $S^2$ thus it does not give
contribution to the anisotropy constant.
The spin factor of the remaining diagram b) is 
\begin{equation}
S^2 F_1(R_n,\theta_n,R_{n'},\theta_{n'})+
S_z^2 F_2(R_n,\theta_n,R_{n'},\theta_{n'}),
\end{equation}
where 
\begin{mathletters}
\begin{equation}
F_1(R_n,\theta_n,R_{n'},\theta_{n'})=
2 J^2 f_{1122},
\end{equation}
\begin{equation}
F_2(R_n,\theta_n,R_{n'},\theta_{n'})=
2 J^2 (f_{1111}-f_{1212}-f_{1122}),
\label{F2}
\end{equation}
\end{mathletters}
with
\begin{equation}
f_{\sigma_1\sigma_2\sigma_3\sigma_4}=\sum\limits_{mm'} 
\tilde W_{mm'\sigma_1\sigma_2}(R_n,\theta_n)
\tilde W_{m'm\sigma_4\sigma_3}(R_{n'},\theta_{n'}).
\end{equation}

After calculation of the remaining part of the diagram Fig.\ \ref{fig3} (b), 
its total contribution is
\begin{equation}
4\rho_0^4Df\left({\omega\over D}\right) 
\bigl (S^2 F_1(R_n,\theta_n,R_{n'},\theta_{n'})+
S_z^2 F_2(R_n,\theta_n,R_{n'},\theta_{n'})\bigr )
\end{equation}
where $\rho_0$ is the density of the states of the conduction electrons for 
one spin direction and $D$ its band width. The function $f$ gives the
analytical part of the diagram which is given in Table~\ref{tab}.

As this diagram contains two host atoms, averages have to be taken over 
$n$ and $n'$ which will be performed in the next Section.

\section{The anisotropy constant}
\label{sec:5}

As it was shown in Section~\ref{sec:4} 
the leading contribution in spin-orbit coupling to the anisotropy 
constant (see Eq.\ (\ref{Haniz})) comes from a second order diagram,
namely from Fig.\ \ref{fig3} (b) 
and it is
\begin{equation}
4\rho_0^4Df\left({\omega\over D}\right) 
F_2(R_n,\theta_n,R_{n'},\theta_{n'}).
\end{equation}
 
\begin{figure}
  \begin{center}
    \epsfig{file=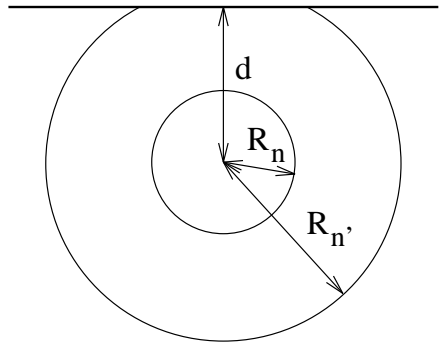}
  \end{center}
\caption{Carrying out the average over the homogeneously dispersed
host atoms, the different shells with constant radiusis $R_n$ and $R_{n'}$ are
shown which can be restricted by the surface.}
\label{fig4}
\end{figure}

As that diagram contains two host atoms with indices $n$
and $n'$, the summation over those must be carried out. According to 
our simple model this gives for the anisotropy factor 
\begin{equation}
K={1\over a^6}\int d^3R_n\int d^3R_{n'} 4\rho_0^4 D f\left({\omega\over
D} \right) F_2(R_n,\theta_n,R_{n'},\theta_{n'}),
\end{equation}
where $a^3$ is the size of the volume per host atom.
The integrations were calculated by considering first the 
shells with constant $R_n$ and $R_{n'}$ (see Fig.\ \ref{fig4}) and
integrating with respect to the angles.
The integration with respect to
$\varphi_n$ and $\varphi_{n'}$ was trivial according to the conservation
of the $z$ component of the angular momentum which was used from the
beginning.

\begin{figure}
  \begin{center}
    \epsfig{file=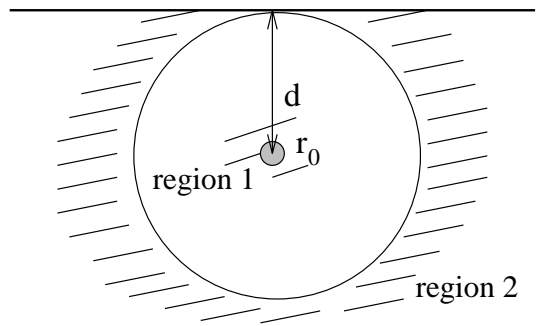}
  \end{center}
\caption{The two dominating regions contributing to the double integral.
Region 2 is formed by those shells of smallest radia which are not complete
due to the presence of the surface. Region 1 is around the impurity of
smallest radia with short range rutoff $r_0$.}
\label{intlimit}
\end{figure}

If e.g. $R_n>d$ then the presence of the surface
appears as a limit in the $\theta_n$ integration, more precisely
we have to
integrate from $\theta_{n,min}=\arccos(d/R_n)$ to $\pi$.
The integrals were calculated in a way
in which the integration regime was divided into four parts where
$R_n, R_{n'}<d$; $R_n>d, R_{n'}<d$; $R_n<d, R_{n'}>d$ and
$R_n, R_{n'}>d$, respectively.
The integrations with respect to $\theta_n$ and $\theta_{n'}$ was
simple and made the contribution of the first part to be zero.
The others give
\begin{eqnarray}
K&=&4\rho_0^4 D f\left ({\omega\over D} \right )\nonumber \\
&&\cdot\biggl ( 
{1\over a^6}\int\limits_d^{\infty} dR_n R_n^2\int\limits_{r_0}^d dR_{n'} 
R_{n'}^2 J_1(R_n,R_{n'})+
{1\over a^6}\int\limits_d^{\infty} dR_n R_n^2\int\limits_d^{\infty} dR_{n'} 
R_{n'}^2 J_2(R_n,R_{n'})\biggr ),
\label{int}\end{eqnarray}
where $r_0$ is a short distance cutoff in range of the atomic radius, and
\begin{mathletters}
\begin{equation}
J_1(R_n,R_{n'})=2\int\limits_{\theta_{nmin}}^{\pi}d\theta_n\sin\theta_n
\int\limits_0^{\pi}d\theta_{n'}\sin\theta_{n'} F(R_n,R_{n'},\theta_n,
\theta_{n'})
\label{J1}\end{equation}
\begin{equation}
J_2(R_n,R_{n'})=\int\limits_{\theta_{nmin}}^{\pi}d\theta_n\sin\theta_n
\int\limits_{\theta_{n'min}}^{\pi}d\theta_{n'}\sin\theta_{n'} 
F(R_n,R_{n'},\theta_n,\theta_{n'}).
\label{J2}
\end{equation}
\end{mathletters}
These remaining integrations with respect to $R_n$ and $R_{n'}$
were estimated in the leading order in $1/(k_F d)$ (see Appendix~\ref{app:2}).
It turned out (see Eq.\ (\ref{integral1}) and (\ref{integral2})) 
that the dominant
contribution arises from the integral of $J_1(R_n,R_{n'})$ where
$R_n>d>R_{n'}$ or the opposite, and the contribution comes from
the lower limits of the integral (see Fig.~\ref{intlimit}),
namely $R_n=d$ and $R_{n'}=r_0$.
Thus for $k_F d\gg1$ 
\begin{equation}
K_d=16 D (J\rho_0)^2 {\Delta^2\lambda^2\over\varepsilon_0^4} f
\left({\omega\over D}\right) {1\over(k_F a)^6} {P(k_F r_0)\over k_F d}>0
\label{Kcon}
\end{equation}
where $P(k_F r_0)$ is a numerical factor depending strongly on $r_0$
(see Eq.\ (\ref{P0})) and it is positive at least for $k_F r_0>0.1$.

\section{Conclusions}
\label{sec:6}

It is shown in the present paper that for a magnetic impurity embodied into 
an infinite electron gas a magnetic anisotropy given by Eq.\ (\ref{Haniz})
is developed if the impurity is surrounded by atoms (e.g. Au, Fe) 
with large spin-orbit interaction in an asymmetrical way. The condition
for formation of that anisotropy is that electrons scattered by the magnetic 
impurities are in angular momentum channels different from zero ($l\neq 0$).
In the other case ($l=0$), the impurity experiences the host atoms
in the same distance from the impurity
in an identical way,
thus the shape of the sample does not play any 
direct role and, therefore, no anisotropy axis can be exhibited.

We considered an infinite half-space with homogeneously dispersed host
atoms with spin-orbit interaction, in which an impurity is placed in a 
distance $d$ from the surface of that half-space. The shape of 
the sample was taken into
consideration only in the position of the host atoms, so the conduction
electrons were assumed to move in the whole space. 
In the calculation no randomness was taken into account, therefore, it is
valid only in the ballistic region.

To describe the interaction between the conduction electrons and the
magnetic impurity we used the simplest realistic Hamiltonian with
orbital quantum numbers. 

For the spin-orbit interaction taking place on the d-levels of the host 
an Anderson like model \cite{Anderson}
is developed with the spin-orbit interaction on the atomic level of 
strength $\lambda$ and the hybridization matrix element $V$. In this
way for the effective spin-orbit interaction between the conduction
electron and the host atom an oversimplified model is obtained
(see Eq.\ (\ref{W})).
The exchange interaction and the spin-orbit
interaction was assumed to be weak, thus perturbation theory was
applied (see Sec.~\ref{sec:2}).

First we calculated the electron propagator in first order of
spin-orbit coupling (see Section~\ref{sec:3}). The angular dependence
was very important because keeping only the s-wave
scattering the spin-orbit interaction cannot influence the impurity
spin dynamics \cite{Meir}.

Then the self-energy corrections for the impurity spin were calculated
by using Abrikosov's pseudofermion representation for the impurity spin
and final temperature Green's function technique for the exchange
interaction (see Sec.~\ref{sec:4}).
It turned out that the first correction to the anisotropy constant
defined in Eq.\ (\ref{Haniz}) comes from a second order diagram in
spin-orbit interaction (see Fig.\ \ref{fig3} (b)), thus an average 
had to be performed 
over the two host atoms (see Sec.~\ref{sec:5} and 
Fig.\ \ref{fig4}\ for the relevant regions of integrations).
The result of this averaging was performed in the leading order in
$1/(k_F d)$ in Appendix~\ref{app:2}, the anisotropy
constant obtained behaves like $K_d\sim 1/(k_F d)$ in the leading order, 
and oscillations occur only in the next order.

This behavior turned out to be
independent of the actual nonzero values of the angular momenta when the 
calculation was repeated 
for different angular momenta of the magnetic impurity ($l'$) and
also of the dominant spin-orbit scattering channel at the host atoms
($l$). Furthermore, in all of the cases $K_d>0$.

To estimate the order of magnitude of the anisotropy factor given by
Eq.\ (\ref{Kcon}), we considered the parameters as $J\rho_0\sim 0.1$
(in the case of relevant Kondo temperature $T_K\sim 0.1 - 1$ K), 
$\lambda\sim 1$ eV, $\Delta\sim D\sim 5$ eV, $\varepsilon_0\sim 2.5$ eV,
$F(\omega/D)\sim 2$ (see Table~\ref{tab}), $k_F a\sim 3$ and
$P(k_F r_0)\sim 10-700$, $k_F r_0\sim 0.3-1.5$ \cite{hiba}.
Thus, the final estimation for its order of magnitude is
\begin{equation}
{0.01\over (d/\text{\AA})}\,\text{eV} < K < {1\over (d/\text{\AA})} 
\,\text{eV}.
\end{equation}

In the present theory for the anisotropy the following approximations
are made:
\begin{description}
\item [(i)] The electrons form an infinite sea and only the 
distribution of the spin-orbit scatterers reflects the "shape" of the 
sample. In a real sample conduction electrons are confined into 
the sample and they can scattered by the surface. In a
mesoscopic sample the surface scattering is rather incoherent because
of the absence of smooth surface, therefore, we expect that 
the qualitative results are not sensitive on the particular model considered.
\item [(ii)] It is assumed that the electrons scattered by the magnetic
impurities do not change their azimuthal quantum number $m$. That 
assumption is valid only for perfectly developed $S=(2 l+1)/2$ spin
($S=5/2$ for $l=2$) \cite{Nozieres,Muhl} (see Sec.~\ref{sec:2} and
the Hamiltonian given by Eq.\ (\ref{H0})). If $S\neq (2 l+1)/2$ then 
the Hund's rule does not ensure that conservation and the
Hamiltonian must contain several terms. We believe, however, that the 
quantitative result including the relative distances 
between levels can be affected 
by that generalization, but the concept of surface anisotropy remains valid.
\item [(iii)] The electrons are treated like free electrons, thus their
elastic mean free path $l_{el}=\infty$. In the reality the elastic mean free
path is finite and the Green's function connecting the impurity and
the spin-orbit scatterers contains an exponential decay. That decay factor 
ensures that the anisotropy is influenced only by those spin-orbit scatterers
which are inside of the region of elastic mean free path.
\end{description}

The geometry treated in the paper is the most simple example. The situation 
is somewhat more complicated e.g. in case of such a thin wire where even
the middle of wire is affected by the anisotropy. Then nearby the surface 
the anisotropy axis is parallel to the normal direction of the surface,
in the middle of the sample the spin direction in the ground state for $S=2$ 
must be, however, perpendicular to all of
the surface elements, thus it must lie along the axis of the wire.
That corresponds to an anisotropy in the direction of wire but with a negative 
coefficient. 

\begin{figure}
  \begin{center}
    \epsfig{file=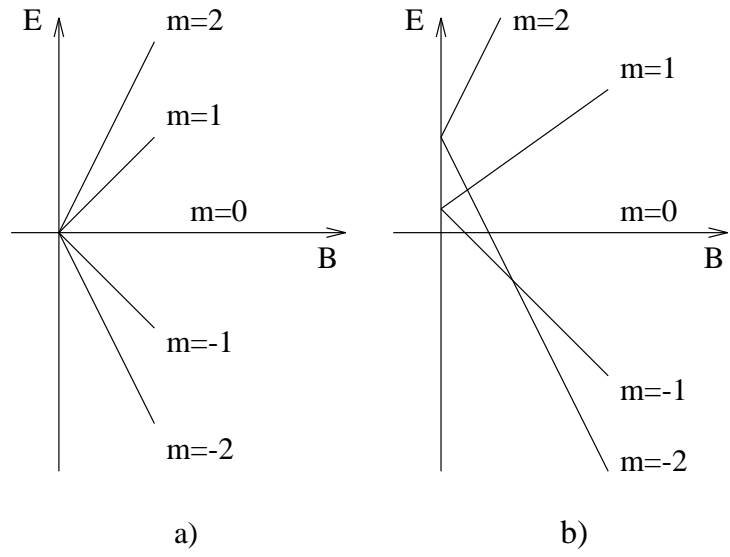}
  \end{center}
\caption{The levels for $S=2$ as the function of the external
magnetic field (a) without and (b) with surface anisotropy.}
\label{figmagres}
\end{figure}

Considering the experimental verification of the surface anisotropy,
there are no direct evidences. Recently Giordano \cite{Giordano}
has performed an experiment which proved difficult to explain with 
previously existing theories and he proposed the presented anisotropy
as a possible proper theoretical explanation. In that work \cite{Giordano}
the magnetoresistance of a thin film is studied well above the Kondo
temperature as a function of the external magnetic field. It was found
that the thin film samples needed larger magnetic field to saturate the
impurity contribution to the resistivity. Considering the surface anisotropy 
in the presence of the field 
${\bf B}$, the Hamiltonian of the magnetic moment
is
\begin{equation}
H=K_d (S^z)^2+B g_B S^z,
\end{equation}
where the field is perpendicular to the film. The levels e.g. for 
$S=2$ as the function of the magnetic field are shown in Fig.\ \ref{figmagres}
without and with surface anisotropy. 

It is clearly shown in Fig.\ \ref{figmagres}  that in the presence of the
anisotropy, because of the level crossing, a larger field $B$ is required 
to separate the lowest energy state
from the other levels in order to saturate the magnetoresistance.
The detailed theory will be published
elsewhere \cite{Borda}.
 
The application of the present theory for resistivity of samples where 
the anisotropy affected regions of impurities are not negligible 
compared to the 
bulk, are the subject of the following paper \cite{Ujsaghy2} (Part II).
In case of $S=2$ at the surface the Kondo effect cannot develop as the spins
are frozen in the state $S^z=0$.

The theory could be developed further in different directions to include the 
surface scattering, to determine $K_d$ in the framework of a realistic atomic
calculation, to consider different geometries, and to take into account the 
elastic mean free path $l_{el}$ in an explicit form. 

The strongest ambiguity in the 
calculation is the short range cutoff $r_0$ appearing in Eq.\ (\ref{P0}).

Finally it is important to emphasize that the present calculation is
beyond the Hartree-Fock approximation (see Sec.~\ref{sec:4}),
thus that anisotropy should not be obtained by band structure calculation
in agreement with the present results obtained by L. Szunyogh {\it et al.}
\cite{Szunyogh}.

\section*{Acknowledgment}

The present authors are grateful for useful discussion with G. Bergmann,
L. Borda, N. Giordano, B. L. Gyorffy, Ph. Nozi\`eres, M. Parpia,
P. Phillips, L. Szunyogh and G. Zar\'and.
The work was supported by grants Hungarian OTKA T02228/1996 and T024005/1997.
One of us (O.\'U.) was supported by TEMPUS Mobility Grant and A.Z. is grateful 
for the support by the Humboldt Foundation.

\appendix

\section{}
\label{app:1}

In this section we derive the restrictions on the electron propagator 
calculated in Section~\ref{sec:3} due to the time-reversal symmetry.
As in \cite{Meir} we calculate  
\begin{eqnarray}
\langle c_{lm\sigma}(r,t) c_{l'm'\sigma'}^{\dagger}(r,0)\rangle&=&
{1\over Z} \sum\limits_a e^{-\beta E_a} \langle\Psi_a | c_{lm\sigma}(r,t)
c_{l'm'\sigma'}^{\dagger}(r,0) | \Psi_a\rangle\nonumber \\
&=&{1\over Z} \sum\limits_{a,b} e^{-\beta E_a} \langle\Psi_a | 
c_{lm\sigma}(r,t) |\Psi_b\rangle\langle \Psi_b| 
c_{l'm'\sigma'}^{\dagger}(r,0) | \Psi_a\rangle\nonumber \\ 
&=&{1\over Z} \sum\limits_{a,b} e^{-\beta E_a+i(E_a-E_b)t}
\langle\Psi_a | c_{lm\sigma}(r,0) |\Psi_b\rangle
\langle \Psi_b| c_{l'm'\sigma'}^{\dagger}(r,0) | \Psi_a\rangle,
\label{mael}\end{eqnarray}
where $c_{lm\sigma}(r,t)$ ($c_{lm\sigma}^{\dagger}(r,t)$)
annihilates (creates) a conduction electron of spin $\sigma$ and orbital
momentum $l,m$ at position $r$ and time $t$.
When the time-reversal symmetry is obeyed 
\begin{eqnarray}
\langle c_{lm\sigma}(r,t) c_{l'm'\sigma'}^{\dagger}(r,0)\rangle&&=
\langle K^+ c_{lm\sigma}(r,t) K K^+ c_{l'm'\sigma'}^{\dagger}(r,0)
K\rangle\nonumber \\
={1\over Z} \sum\limits_{a,b}&&e^{-\beta E_a+i(E_a-E_b)t}
\langle\Psi_a | K^+ c_{lm\sigma}(r,0) K|\Psi_b\rangle
\langle \Psi_b| K^+ c_{l'm'\sigma'}^{\dagger}(r,0) K|\Psi_a\rangle
\nonumber \\
={1\over Z} \sum\limits_{a,b}&&e^{-\beta E_a+i(E_a-E_b)t}
(-1)^{1/2+\sigma+l+m}\langle \Psi_b| c_{l,-m,-\sigma}^{\dagger}(r,0)
|\Psi_a\rangle \nonumber \\
&&\cdot (-1)^{1/2+\sigma'+l'+m'}
\langle\Psi_a | c_{l',-m',-\sigma'}(r,0) |\Psi_b\rangle,
\end{eqnarray}
where $K$ is the time-reversal operator.

In comparison with Eq.\ (\ref{mael}) we obtain the relation
\begin{equation}
\langle c_{lm\sigma}(r,t) c_{l'm'\sigma'}^{\dagger}(r,0)\rangle=
(-1)^{1+\sigma+\sigma'+l+l'+m+m'}
\langle c_{l',-m',-\sigma'}(r,t) c_{l,-m,-\sigma}^{\dagger}(r,0)\rangle.
\label{rel1}\end{equation}
Applying the same procedure to $\langle c_{l'm'\sigma'}^{\dagger}(r,0)
c_{lm\sigma}(r,t) \rangle$ the obtained relation is
\begin{equation}
\langle c_{l'm'\sigma'}^{\dagger}(r,0) c_{lm\sigma}(r,t)\rangle=
(-1)^{1+\sigma+\sigma'+l+l'+m+m'}
\langle c_{l,-m,-\sigma}^{\dagger}(r,0) c_{l',-m',-\sigma'}(r,t) \rangle.
\label{rel2}\end{equation}
Thus in the case of s-wave scattering ($l=l'=m=m'=0$) 
the electron propagator is diagonal in spin space in agreement
with \cite{Meir}.
In our case ($l=l'=2$) the restrictions for the 
electron propagator given by the time-reversal symmetry 
(see Eq.\ (\ref{rel1}) and Eq.\ (\ref{rel2})) are
\begin{equation}
\tilde W_{mm'\sigma\sigma'}(R_n,\theta_n,\varphi_n)=(-1)^{1+\sigma+
\sigma'+m+m'} \tilde W_{-m',-m,\sigma',-\sigma}(R_n,\theta_n,\varphi_n),
\end{equation}
which served a good check for the calculation.

\section{}
\label{app:2}

Here we estimate the integrals 
\begin{mathletters}
\begin{equation}
I_1={1\over a^6}\int\limits_d^{\infty} dR_n R_n^2\int\limits_{r_0}^d 
dR_{n'} R_{n'}^2 J_1(R_n,R_{n'})
\end{equation}
\begin{equation}
I_2={1\over a^6}\int\limits_d^{\infty} dR_n R_n^2\int\limits_d^{\infty}
dR_{n'} R_{n'}^2 J_2(R_n,R_{n'})
\end{equation}
\end{mathletters}
appearing in Eq.\ (\ref{int}), in leading order
in $1/(k_F d)$ for $k_F d\gg1$.
This calculation is very long for $I_2$, but similar to $I_1$, thus
we present here the estimation only for $I_1$, but at the end we give the
form for $I_2$, too.

After the integration with respect to $\theta_n$ and $\theta_{n'}$ 
\begin{eqnarray}
J_1(R_n,R_{n'})&&=
{16\over 15} J^2 \biggl ({2\pi V^2\lambda\over\varepsilon_0^2}\biggr )^2
{d (R_n^2-d^2)\over R_n^3}\nonumber \\
&&\cdot\left ( 3 v_{k_F0}(R_n) v_{k_F1}(R_n)
-v_{k_F1}^2(R_n)+2 v_{k_F1}(R_n) v_{k_F2}(R_n)-4 v_{k_F2}^2(R_n)\right )
\nonumber \\
&&\cdot\left (6 v_{k_F0}(R_{n'}) v_{k_F1}(R_{n'})
+v_{k_F1}^2(R_{n'})+4 v_{k_F1}(R_{n'}) v_{k_F2}(R_{n'})
+4 v_{k_F2}^2(R_{n'})\right ).
\end{eqnarray}
Substituting the $v_{k_Fm}(R_n)$ matrix element from Eq.\ (\ref{vkm}) 
into the integral, using trigonometric identities
and introducing the dimensionless integration variables 
$s=k_F R_n$, $t=k_F R_{n'}$ and notations $x=k_F d$, $x_0=k_F r_0$,
the integral has the form
\begin{eqnarray}
I_1&=&{16\over 15} J^2 \bigl ({2\pi V^2\lambda\over\varepsilon_0^2}\bigr )^2 
\bigl ({900\over k_F^3}\bigr )^2 x {1\over a^6}\nonumber \\ 
&&\cdot\int\limits_x^{\infty} ds (s^2-x^2) \biggl (
{225\over 2 s^{11}}-{225\cos 2 s\over 2 s^{11}}-{225\sin 2 s\over s^{10}}
+{18\over s^9}+{207\cos 2 s\over s^9}+{114\sin 2 s\over s^8}\nonumber \\
&+&{15\over 8 s^7}-{327\cos 2 s\over 8 s^7}-{39\sin 2 s\over 4 s^6}
+{1\over 4 s^5}+{3\cos 2 s\over 2 s^5}+{1\sin 2 s\over 8 s^4}\biggr )
\nonumber \\
&&\cdot\int\limits_{x_0}^x dt \biggl (
{315\over 2 t^8}-{315\cos 2 t\over 2 t^8}-{315\sin 2 t\over t^7}
+{45\over 2 t^6}+{585\cos 2 t\over 2 t^6}+{165\sin 2 t\over t^5}\nonumber \\
&+&{15\over 8 t^4}-{495\cos 2 t\over 8 t^4}-{63\sin 2 t\over 4 t^3}
+{1\over 8 t^2}+{21\cos 2 t\over 8 t^2}+{1\sin 2 t\over 4 t}\biggr ).
\end{eqnarray}
The occuring integrals are the type of
\begin{mathletters}
\label{type1}
\begin{equation}
\int\limits_\alpha^\beta ds (s^2-x^2) {\sin 2 s\over s^n}
\end{equation}
\begin{equation}
\int\limits_\alpha^\beta ds (s^2-x^2) {\cos 2 s\over s^n}
\end{equation}
\begin{equation}
\int\limits_\alpha^\beta ds (s^2-x^2) {1\over s^n}
\end{equation}
\end{mathletters}
and
\begin{mathletters}
\label{vizsint}
\begin{equation}
\int\limits_\alpha^\beta dt {\sin 2 t\over t^n}
\end{equation}
\begin{equation}
\int\limits_\alpha^\beta dt {\cos 2 t\over t^n}
\end{equation}
\begin{equation}
\int\limits_\alpha^\beta dt {1\over t^n}={1\over n-1} {1\over \alpha^{n-1}}
-{1\over n-1} {1\over \beta^{n-1}}.
\end{equation}
\end{mathletters}
Let us consider the first two integrals of the second type.
After integration by part they are \cite{Ryzhik}
\begin{eqnarray}
\int dt {\sin 2 t\over t^{2 k}}&=&\varphi_s(2 k,2 t)=
(-1)^{k+1} {2^{2 k-1}\over (2 k-1)!}
\biggl \{ 
\sum\limits_{i=0}^{k-1} (-1)^{i+1} (2 i)! {\sin 2 t\over (2 t)^{2 i+1}}
\nonumber \\   
&+&\sum\limits_{i=0}^{k-2} (-1)^{i} 
(2 i+1)! {\cos 2 t\over (2 t)^{2 i+2}}\biggr \}\nonumber \\
&+&(-1)^{k+1} {2^{2 k-1}\over (2 k-1)!} Ci\left [2 t\right ],
\label{reks1}
\end{eqnarray}
\begin{eqnarray}
\int dt {\sin 2 t\over t^{2 k+1}}&=&\varphi_s(2 k+1,2 t)=
(-1)^{k+1} {2^{2 k}\over (2 k)!}
\biggl \{ 
\sum\limits_{i=0}^{k-1} (-1)^{i+1}
(2 i+1)! {\sin 2 t\over (2 t)^{2 i+2}}
\nonumber \\   
&+&\sum\limits_{i=0}^{k-1} (-1)^{i+1} 
(2 i)! {\cos 2 t\over (2 t)^{2 i+1}}\biggr \} \nonumber \\
&+&(-1)^{k} {2^{2 k}\over (2 k)!} Si\left [2 t\right ]
\label{reks2}\end{eqnarray}
and
\begin{eqnarray}
\int\limits dt {\cos 2 t\over t^{2 k}}&=&\varphi_c(2 k,2 t)=
(-1)^{k+1} {2^{2 k-1}\over (2 k-1)!}
\biggl \{ 
\sum\limits_{i=0}^{k-1} (-1)^{i+1}
(2 i)! {\cos 2 t\over (2 t)^{2 i+1}}
\nonumber \\   
&-&\sum\limits_{i=0}^{k-2} (-1)^{i}
(2 i+1)! {\sin 2 t\over (2 t)^{2 i+2}}\biggr \}\nonumber \\
&+&(-1)^{k} {2^{2 k-1}\over (2 k-1)!} Si\left [2 t\right ],
\label{rekc1}
\end{eqnarray}
\begin{eqnarray}
\int\limits dt {\cos 2 t\over t^{2 k+1}}&=&\varphi_c(2 k+1,2 t)=
(-1)^{k+1} {2^{2 k}\over (2 k)!}
\biggl \{ 
\sum\limits_{i=0}^{k-1} (-1)^{i+1}
(2 i+1)! {\cos 2 t\over (2 t)^{2 i+2}}
\nonumber \\   
&-&\sum\limits_{i=0}^{k-1} (-1)^{i+1}
(2 i)! {\sin 2 t\over (2 t)^{2 i+1}}\biggr \}\nonumber \\
&+&(-1)^{k} {2^{2 k}\over (2 k-1)!} Ci\left [2 t\right ]
\label{rekc2}\end{eqnarray}
where
\begin{mathletters}
\begin{equation}
Ci\left [t\right ]=-\int\limits_t^{\infty} du {\cos u\over u}
\end{equation}
\begin{equation}
Si\left [t\right ]=\int\limits_0^t du {\sin u\over u}
\end{equation}
\end{mathletters}
are the Cosine and Sine Integral functions \cite{Ryzhik}.
To estimate the integrals in Eq.\ (\ref{vizsint}) for $\alpha$ or 
$\beta\gg1$ the 
expression of the Cosine and Sine Integral function in terms of 
auxiliary functions was used \cite{Abram}
\begin{mathletters}
\begin{equation}
Ci\left [t\right ]=f(t)\sin t-g(t)\cos t
\end{equation}
\begin{equation}
Si\left [t\right ]={\pi\over 2}-f(t)\cos t-g(t)\sin(t),
\end{equation}
\end{mathletters}
where
\begin{mathletters}
\begin{equation}
f(t)\sim {1\over t}\bigl (1-{2!\over t^2}+{4!\over t^4}-{6!\over t^6}+\dots
\bigr )={1\over t}\sum\limits_{i=0}^{\infty} (-1)^i {(2 i)!\over t^{2 i}}
\end{equation}
\begin{equation}
g(t)\sim {1\over t^2}\bigl (1-{3!\over t^2}+{5!\over t^4}-{7!\over t^6}+\dots
\bigr )={1\over t^2}\sum\limits_{i=0}^{\infty} (-1)^i {(2 i+1)!\over t^{2 i}}.
\end{equation}
\end{mathletters}
It can be seen from these asymptotic expansions that the primitive
functions in 
Eq.\ (\ref{vizsint}) are e.g. for $\alpha\gg1$
\begin{eqnarray}
\varphi_s(2 k,2 \alpha)&\sim&(-1)^{k-1} 
{2^{2 k-1}\over (2 k-1)!} \biggl [
{\sin 2 \alpha\over 2 \alpha}\sum\limits_{i=k}^{\infty} (-1)^i 
{(2 i)!\over (2 \alpha)^{2 i}}\nonumber \\
&-&{\cos 2 \alpha\over (2 \alpha)^2}\sum\limits_{i=k-1}^{\infty} (-1)^i 
{(2 i+1)!\over (2 \alpha)^{2 i}}\biggr ],
\end{eqnarray}
\begin{eqnarray}
\varphi_s(2 k+1,2 \alpha)&\sim&(-1)^{k} {2^{2 k}\over (2 k)!} \biggl [
{\pi\over 2}-{\cos 2 \alpha\over 2 \alpha}\sum\limits_{i=k}^{\infty} (-1)^i 
{(2 i)!\over (2 \alpha)^{2i}}\nonumber \\
&-&{\sin 2 \alpha\over (2 \alpha)^2}\sum\limits_{i=k}^{\infty} (-1)^i
\end{eqnarray}
and
\begin{eqnarray}
\varphi_c(2 k,2 \alpha)&\sim&(-1)^{k} 
{2^{2 k-1}\over (2 k-1)!} \biggl [
{\pi\over 2}-{\cos 2 \alpha\over 2 \alpha}\sum\limits_{i=k}^{\infty} (-1)^i 
{(2 i)!\over (2 \alpha)^{2i}}\nonumber \\
&-&{\sin 2 \alpha\over (2 \alpha)^2}\sum\limits_{i=k-1}^{\infty} (-1)^i
{(2 i+1)!\over (2 \alpha)^{2 i}}\biggr ],
\end{eqnarray}
\begin{eqnarray}
\varphi_c(2 k+1,2 \alpha)&\sim&(-1)^{k} {2^{2 k}\over (2 k)!} \biggl [
{\sin 2 \alpha\over 2 \alpha}\sum\limits_{i=k}^{\infty} (-1)^i {(2 i)!\over (2 \alpha)^{2i}}
\nonumber \\
&-&{\cos 2 \alpha\over (2 \alpha)^2}\sum\limits_{i=k}^{\infty} (-1)^i
{(2 i+1)!\over (2 \alpha)^{2 i}}\biggr ].
\end{eqnarray}
Thus in our case when $\alpha=x\gg1$, $\beta=\infty$ or 
$\alpha=x_0$, $\beta=x\gg1$ 
the contributions of these integrals in leading order in $1/x$ are
\begin{mathletters}
\label{lim}
\begin{equation}
\int\limits_x^{\infty} dt {\sin 2 t\over t^n}\sim {\cos 2 x\over 2 x^n},
\end{equation}
\begin{equation}
\int\limits_x^{\infty} dt {\cos 2 t\over t^n}\sim -{\sin(2 x)\over 2 x^n},
\end{equation}
\begin{equation}
\int\limits_{x_0}^x dt {\sin 2 t\over t^n}\sim -\varphi_s(n,x_0),
\end{equation}
\begin{equation}
\int\limits_{x_0}^x dt {\cos 2 t\over t^n}\sim -\varphi_c(n,x_0),
\end{equation}
\end{mathletters}
where $\varphi_s(n,x_0)$ and $\varphi_c(n,x_0)$ denote the primitive
functions of the integrals given in Eq.\ (\ref{reks1}), (\ref{reks2}),
(\ref{rekc1}), (\ref{rekc2}).

Turning to the integrals of the first type in Eq.\ (\ref{type1}) they can be
transformed by integration by part into
\begin{mathletters}
\begin{equation}
\int\limits_\alpha^\beta ds (s^2-x^2) {\sin 2 s\over s^n}=\biggl [{s^2-x^2\over s^n} 
(-{\cos 2 s\over 2})\biggr ]_\alpha^\beta+\int\limits_\alpha^\beta ds {\cos 2 s\over s^{n-1}}
-\int\limits_\alpha^\beta ds {n\over 2} (s^2-x^2) {\cos 2 s\over s^{n+1}},
\end{equation}
\begin{equation}
\int\limits_\alpha^\beta ds (s^2-x^2) {\cos 2 s\over s^n}=\biggl [{s^2-x^2\over s^n}
{\sin 2 s\over 2} \biggr ]_\alpha^\beta -\int\limits_\alpha^\beta ds {\sin 2 s\over s^{n-1}}
+\int\limits_\alpha^\beta ds {n\over 2} (s^2-x^2) {\sin 2 s\over s^{n+1}},
\end{equation}
\begin{equation}
\int\limits_\alpha^\beta ds (s^2-x^2) {1\over s^n}=\biggl [{1\over (3-n) s^{n-3}}
-{x^2\over (1-n) s^{n-1}}\biggr ]_\alpha^\beta.
\end{equation}
\end{mathletters} 
Using the leading order formulas for the integrals of the second type 
in Eq.\ (\ref{lim}) and considering our case ($\alpha=x\gg1$, $\beta=\infty$) 
the integrals above in the leading order in $1/x$ are
\begin{mathletters}
\begin{equation}
\int\limits_x^{\infty} ds (s^2-x^2) {\sin 2 s\over s^n}\sim
{\sin 2 x\over 2 x^{n-1}},
\end{equation}
\begin{equation}
\int\limits_x^{\infty} ds (s^2-x^2) {\cos 2 s\over s^n}\sim
{\cos 2 x\over 2 x^{n-1}},
\end{equation}
\begin{equation}
\int\limits_x^{\infty} ds (s^2-x^2) {1\over s^n}\sim
\biggl ({1\over (n-3)}-{1\over (n-1)}\biggr ){1\over x^{n-3}}.
\end{equation}
\end{mathletters}

Thus the final estimation for the $I_1$ 
integral in the leading order in $1/x$ ($x=k_F d$) is
\begin{equation}
I_1=4 J^2 \biggl({\pi V^2\lambda\over\varepsilon_0^2}\biggr)^2 
{1\over (k_F a)^6}
\biggl [{P(x_0) \over x}-{P(x_0) \sin(2 x)\over x^2}-{6750 (1+\cos(2 x))
\over x^2}\biggr ]
\label{integral1}\end{equation}
where
\begin{eqnarray}
P(x_0)&=&{54000}\biggl ({315\over 2}\varphi_c(8,x_0)
-{585\over 2}\varphi_c(6,x_0)
+{495\over 8}\varphi_c(4,x_0)-{21\over 8}\varphi_c(2,x_0)\nonumber \\
&+&315\varphi_s(7,x_0)-165\varphi_s(5,x_0)
+{63\over 4}\varphi_s(3,x_0)-{1\over 4}\varphi_s(1,x_0)\nonumber \\
&+&{45\over 2 x_0^7}+{9\over 2 x_0^5}+{5\over 8 x_0^3}
+{1\over 8 x_0}\biggr )\nonumber \\
&=&{54000}\biggl (
{{45}\over {2\,{x_0^7}}} + {9\over {2\,{x_0^5}}} + {5\over {8\,{x_0^3}}} + 
  {1\over {8\,x_0}} - {{45\,\cos (2\,x_0)}\over {2\,{x_0^7}}}\nonumber \\ 
 &+&{{81\,\cos (2\,x_0)}\over {2\,{x_0^5}}} - {{53\,\cos (2\,x_0)}\over
{8\,{x_0^3}}} +{{\cos (2\,x_0)}\over {8\,x_0}} - {{45\,\sin (2\,x_0)}\over
{{x_0^6}}} \nonumber \\
&+& {{21\,\sin (2\,x_0)}\over {{x_0^4}}} - {{5\,\sin (2\,x_0)}\over
{4\,{x_0^2}}}\biggr ).
\label{P0}\end{eqnarray}

The $I_2$ integral in leading order in $1/x$ ($x=k_F d$) is
\begin{equation}
I_2=4 J^2 \biggl({\pi V^2\lambda\over\varepsilon_0^2}\biggr)^2 
{1\over (k_F a)^6}
\biggl [{6750\cos(2 x)\over x^2}+{347625\over 64 x^2}\biggr ].
\label{integral2}\end{equation}

\begin{table}
\begin{center}
    \epsfig{file=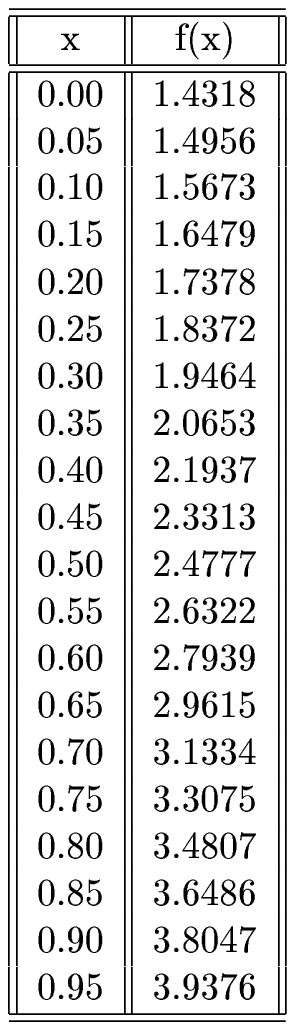}
\end{center}
\caption{The analytical part of diagram Fig.~\ref{fig4} b) 
in function of $x={\omega\over D}$.}
\label{tab}
\end{table}


\begin{thebibliography}{99}
\bibitem{hiba} The error in the estimation of the numerical factor $P(x_0)$
in \cite{Ujsaghy,proc} was corrected here.
\end{thebibliography}
\end{document}